\documentclass[%
 aip,
 amsmath,amssymb,
reprint,%
]{revtex4-1}

\usepackage{eso-pic, rotating,graphicx}
\usepackage{dcolumn}
\usepackage{bm}

\usepackage[utf8]{inputenc}
\usepackage[T1]{fontenc}
\usepackage{mathptmx}
\usepackage{etoolbox}

\usepackage{siunitx}
\usepackage{hyperref}
\hypersetup{colorlinks =true,linkcolor = [rgb]{0.0,0.0,1.0},citecolor =[rgb]{0.0,0.0,1.0},filecolor =[rgb]{0.0,0.0,1.0},urlcolor = [rgb]{0.0,0.0,1.0}}
\usepackage[nameinlink]{cleveref}
\crefname{figure}{Fig.}{Figs.}
\crefname{equation}{Eq.}{Eqs.}
\crefname{table}{Table}{Tables}
\crefname{section}{Section}{Sections}
\Crefname{figure}{Fig.}{Figs.}
\Crefname{equation}{Eq.}{Eqs.}
\Crefname{table}{Table}{Tables}
\Crefname{section}{Section}{Sections}

\makeatletter
\def\@email#1#2{%
 \endgroup
 \patchcmd{\titleblock@produce}
  {\frontmatter@RRAPformat}
  {\frontmatter@RRAPformat{\produce@RRAP{*#1\href{mailto:#2}{#2}}}\frontmatter@RRAPformat}
  {}{}
}%

\AddToShipoutPicture{\put(1.5cm,1.5cm){\rotatebox{0}{\scalebox{1.0}{\textcolor{blue!50}{This article may be downloaded for personal use only. Any other use requires prior permission of the author and AIP Publishing.}}}}}
\AddToShipoutPicture{\put(1.2cm,1.0cm){\rotatebox{0}{\scalebox{1.0}{\textcolor{blue!50}{This article appeared in A. R. Peeketi et al., J. Chem. Phys. 160, 104902 (2024) and may be found at https://doi.org/10.1063/5.0187320}}}}}
\AddToShipoutPicture{\put(20.1cm,4cm){\rotatebox{90}{\scalebox{1.1}{\textcolor{red!50}{This is the accepted manuscript version of the article published in The Journal of Chemical Physics}}}}}
\AddToShipoutPicture{\put(20.8cm,4.0cm){\rotatebox{90}{\scalebox{1.1}{\textcolor{blue!50}{Please cite the article as A. R. Peeketi et al., J. Chem. Phys. 160, 104902 (2024), DOI:\href{https://doi.org/10.1063/5.0187320}{https://doi.org/10.1063/5.0187320}}}}}}

\newcommand{\update}[1]{{\textcolor{black}{#1}}}

\newcommand{\trans}{{\emph{trans}}}
\newcommand{\cis}{{\emph{cis}}}

\newcommand{\nc}{n_c}
\newcommand{\nt}{n_t}
\newcommand{\Ptc}{P_{t \to c}}
\newcommand{\Pct}{P_{c \to t}}
\newcommand{\ntc}{n_{t \to c}}
\newcommand{\nct}{n_{c \to t}}
\newcommand{\ncc}{n_{c \to c}}

\newcommand{\Iuv}{I_{\text{365}}}
\newcommand{\Ivi}{I_{\text{455}}}

\newcommand{\Etc}{E^d_{t \to c}}
\newcommand{\Ect}{E^d_{c \to t}}
\newcommand{\Eth}{E^d_{\text{thermal}}}
\newcommand{\Ed}{E^d_{\alpha}}

\newcommand{\ts}{\tau_{s}}
\newcommand{\Phicnnc}{\phi}
\newcommand{\Phicri}{\phi_{\text{cr}}}

\newcommand{\unitsI}{\milli\watt\per\square\centi\meter}

\newcommand{\ps}{\pico\second}
\newcommand{\ns}{\nano\second}
\newcommand{\nm}{\nano\meter}
\newcommand{\fs}{\femto\second}
\DeclareSIUnit{\calorie}{cal}

\makeatother
\begin{document}
\preprint{JCP23-AR-04008}

\title[]{Photo-activated dynamic isomerization induced large density changes in liquid crystal polymers: A molecular dynamics study}
\author{Akhil Reddy Peeketi}
\affiliation{Center for Soft and Biological Matter, Indian Institute of Technology Madras, Chennai - 600036, India}
\affiliation{Department of Mechanical Engineering, Indian Institute of Technology Madras, Chennai - 600036, India}
\author{Edwin Joseph}
\affiliation{Department of Mechanical Engineering, Indian Institute of Technology Madras, Chennai - 600036, India}
\author{Narasimhan Swaminathan}
\altaffiliation{Corresponding author: Email: n.swaminathan@iitm.ac.in}
\affiliation{Center for Soft and Biological Matter, Indian Institute of Technology Madras, Chennai - 600036, India}
\affiliation
{Department of Mechanical Engineering, Indian Institute of Technology Madras, Chennai - 600036, India}
\author{Ratna Kumar Annabattula}
\altaffiliation{Corresponding author: Email: ratna@iitm.ac.in}
\affiliation{Center for Soft and Biological Matter, Indian Institute of Technology Madras, Chennai - 600036, India}
\affiliation{Department of Mechanical Engineering, Indian Institute of Technology Madras, Chennai - 600036, India}


\date{\today}

\begin{abstract}
We use molecular dynamics simulations to unravel the physics underpinning the light-induced density changes caused by the dynamic \trans-\cis-\trans~isomerization cycles of azo-mesogens embedded in a liquid crystal polymer network, an intriguing experimental observation reported in the literature. We employ two approaches, cyclic and probabilistic switching of isomers, to simulate dynamic isomerization. The cyclic switching of isomers confirms that dynamic isomerization can lead to density changes at specific switch-time intervals. The probabilistic switching approach further deciphers the physics behind the non-monotonous relation between density reduction and light intensities observed in experiments. Light intensity variations in experiments are accounted for in simulations by varying the \trans-to-\cis~and \cis-to-\trans~isomerization probabilities. The simulations show that an optimal combination of these two probabilities results in a maximum density reduction corroborating the experimental observations. At such an optimal combination of probabilities, the dynamic \trans-\cis-\trans~isomerization cycles occur at a specific frequency causing significant distortion in the polymer network, resulting in a maximum density reduction. 
\end{abstract}

\maketitle


\section{\label{Sec:Introduction}Introduction}
\begin{figure*}[!htb]
    \centering
    \includegraphics[width=17.80 cm]{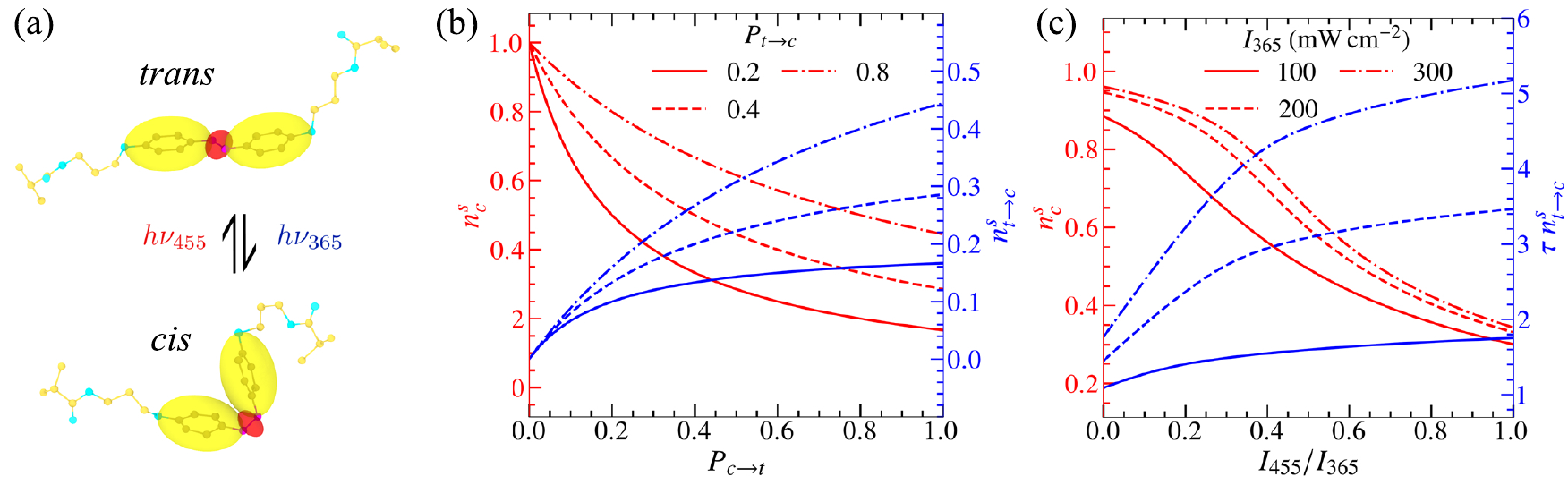}
    \caption{(a) Schematic illustration of \trans~and \cis~states of the azobenzene mesogen. (b) Steady-state mass-fraction of \cis~(left ordinate) and mass-fraction of isomers that are isomerizing from \trans~state to \cis~(right ordinate) for different probabilities of \trans-to-\cis~and \cis-to-\trans~isomerizations. (c) Steady-state mass-fraction of \cis~(left ordinate) and mass-fraction of isomers that are isomerizing from \trans~to \cis~state (right ordinate) for different intensities of \SI{365}{\nm} and \SI{455}{\nm} wavelength light sources following the photo-physical model.\cite{liu2017enhanced}}
    \label{fig:Figure_NC_NTC}
\end{figure*}

Liquid crystal polymers are one of the highly sought-after materials for developing soft actuators\cite{da2020bioinspired,jayoti2023geometry,he2019bioinspired,dradrach2023light}. The ability to program the alignment in liquid crystal polymers allows complex and reversible shape deformations such as curls, twists and cones.\cite{mehta2020design,white2015programmable,peeketi2022calla,ware2015voxelated} Incorporating photo-responsive moieties such as azobenzenes, spiropyrans, nano-particles etc., makes the otherwise thermo-responsive liquid crystal polymers photo-responsive.\cite{bisoyi2016light, huang2021bioinspired,jayoti2022dynamics,wu2011nir,li2017polydopamine,cheng2015nir} A major advantage of photo-responsive systems is non-contact remote actuation, leading to their precise spatial and temporal control.\cite{stoychev2019light,chen2021light} Azobenzene-modified liquid crystal polymer networks (ALCNs) have been explored extensively for applications in soft robotics, photonics, haptics and microfluidics.\cite{ikeda2007photomechanics,liu2013liquid,pang2019photodeformable,mehta2020design} The recent rise in the applications of ALCNs necessitates the need to understand the mechanisms underpinning the light-responsive actuation in greater detail. The earlier consensus of the scientific community was that the bending of the azobenzene as it isomerizes from the straight rod-like \trans~state to bent shaped \cis~state (see \cref{fig:Figure_NC_NTC}a) upon exposure to light creates a network pulling effect causing changes in the local orientational order of the liquid crystal mesogens resulting in macroscopic mechanical deformations (contraction in the direction parallel to the average orientation of mesogens and expansion in the other two perpendicular directions).\cite{garcia2009nematic, sanchez2011opto, da2019unravelling} The magnitude of the contraction/expansion was believed to be proportional to the average mass-fraction of the \cis~isomers.\cite{Corbett2008, Cheng2012,liu2018topographical,korner2020nonlinear,mehta2020modeling,peeketi2021modeling} However, recent works suggest a new mode of generating deformations in liquid crystal polymers.\cite{liu2015new,cheng2017photomechanical} In particular, these studies demonstrate that continuous, dynamic \trans-\cis-\trans~isomerizations collectively generate significant deformations far greater than those caused solely by the statistical accumulation of \cis~isomers.\cite{liu2015new} Furthermore, there is a considerable reduction in the density of the polymer due to the free volume generated by the azo-molecules undergoing dynamic \trans-\cis-\trans~isomerization cycles.\cite{liu2015new,cheng2017photomechanical} \update{Additionally, it is important to note that only 2 wt\% of azo-molecules within the entire ALCN caused a significant 12\% reduction in density.\cite{liu2015new}}

The azo-mesogens usually occur in the stable \trans~state and upon illuminating with Ultra-Violet wavelength (\SI{365}{\nm}), the \trans~isomers isomerize to a meta-stable \cis~state. The \cis~isomers revert to \trans~state when illuminated with Visible wavelength (\SI{455}{\nm}). Even without any illumination, the \cis~isomers still revert to \trans~although slowly through thermal back relaxation. It should be noted that, illuminating with only \SI{365}{\nm} also promotes \cis-to-\trans~isomerization, the extent of the isomerization however pales in comparison to \SI{365}{\nm} promoting \trans-to-\cis~isomerization. However, when the azo-mesogens are illuminated with both \SI{365}{\nm} and \SI{455}{\nm} illuminations, both the \trans-to-\cis~and \cis-to-\trans~reactions are promoted resulting in continuous dynamic \trans-\cis-\trans~isomerization cycles. If the density reduction (or volume expansion) shown by ALCNs\cite{liu2015new} is due to such dynamic isomerizations under the influence of a combined UV and Visible illumination, then the density change should continuously increase with intensities~($I$) of the two light sources~($\Iuv$ and $\Ivi$). To elucidate the aforementioned point, consider the following simplified probabilistic approach. Let a fully \trans~system be subjected to dual wavelength (\SI{365}{\nm} and \SI{455}{\nm}) illumination of certain intensities. Such a system can show both \trans-to-\cis~and \cis-to-\trans~isomerizations with probabilities of $\Ptc$ and $\Pct$, respectively. Then the fraction of \cis~isomers at step $i+1$ ($\nc^{i+1}$) is the sum of the fractions that isomerized from \trans~state at $i$ ($\ntc^i$) and the isomers that remained in \cis~state from $i$ ($\ncc^i$) leading to
\begin{equation}
    \nc^{i+1} = \ntc^i + \ncc^i = \nt^{i} \Ptc + \nc^{i} (1 - \Pct).
\end{equation}
Such an evolution of isomers leads to a steady state after $s+1$ steps when $\nc^{s+1} = \nc^{s}$. The steady state values are given by,
\begin{equation}
    \nc^{s+1} = \nc^{s} = \frac{\Ptc}{\Ptc + \Pct}, \, \nt^{s+1} = \nt^{s} = \frac{\Pct}{\Ptc + \Pct}.
\end{equation}
However, note that some isomers still isomerize from \trans~ state to \cis, while some from \cis~ state to \trans~at such a steady state resulting in a dynamic equilibrium. At this dynamic equilibrium, the fraction of molecules that are isomerizing from \trans~state to \cis~and from \cis~state to \trans~are equal and given by,
\begin{equation}
    \ntc^{s} = \nct^{s} = \frac{\Ptc \Pct}{\Ptc + \Pct}.
\end{equation}
Assuming that the probabilities are directly proportional to the intensities of the corresponding wavelengths (i.e., $\Ptc$ $\propto$ $\Iuv$ and $\Pct$ $\propto$ $\Ivi$), $\nc^{s}$ decreases with increase in $\Pct$ for a given $\Ptc$ (see \cref{fig:Figure_NC_NTC}b). \update{However, the fraction of isomers undergoing dynamic isomerization at steady-state ($\nct^s$) increases with increasing $\Ivi$.} Even with the consideration of full isomerization kinetics and corresponding absorption coefficients of the isomers,\cite{liu2017enhanced} $\nc^s$ decreases with increase in $\Ivi$/$\Iuv$, while $\nct^s$ increases as shown in \cref{fig:Figure_NC_NTC}c (see Supplementary Note S1 for details). Since the dynamic \trans-\cis-\trans~isomerization cycles increase with $\Ivi$, a more significant reduction in density is expected for higher values of $\Ivi$. However, the experimental observations\cite{liu2015new} suggest \update{the existence of a maxima for the density decrease at a specific intensity ratio} (see Fig. S1). Specifically, the density reduction maximizes when $\Ivi$ is almost 10\% of $\Iuv$. That is, one must expect that at higher intensities of both UV and Visible illuminations, \trans-to-\cis~and \cis-to-\trans~reactions are both promoted leading to more frequent \trans-\cis-\trans~conversions. \update{Such a situation must result in larger density decrease with increase in intensities. However, it is observed that increasing $\Ivi$ beyond a certain limit reduced the decrease in density and the decrease dropped to just 2\% at $\Ivi$ = $\Iuv$. Consequently, the experiments indicate a non-monotonous density reduction with increasing intensity ratios. The constitutive equations proposed by \citet{liu2017enhanced} captures this effect through a phenomenological model and ascribes the non-monotonous relationship between density decrease and intensity ratios to the interplay between the oscillating azobenzenes and the distortion of the viscoelastic polymer network. However, the molecular underpinnings on the origins of such a behaviour has still not been explored.} 

Experimental quantification of the dynamic isomerization and its influence on the polymer network is not feasible as it requires tracking individual mesogens inside the network with a very fine spatial and temporal resolution. However, molecular simulations can bridge this gap by accounting for individual atoms building the ALCN and simulating the distortion of the polymer network as the azo-molecules isomerize. Such simulations can reveal the underlying physics of photo-induced density reduction. The existing computational studies at the molecular level analyzed the contribution of static \trans-to-\cis~isomerization\cite{choi2014photo,choi2016molecular,moon2019multiscale} and photo-induced reorientation\cite{ilnytskyi2008molecular,ilnytskyi2011opposite} of the azo-mesogens to the deformation of the ALCN unit cell. Moreover, the computational samples in most of these studies consist of 100\% azo-mesogens.\cite{ilnytskyi2008molecular,ilnytskyi2011opposite,choi2014photo,choi2016molecular,moon2019multiscale} There have been no computational studies at a molecular level to confirm the contribution of dynamic \trans-\cis-\trans~isomerization cycles to density changes of the ALCN, when the azo-mesogen concentration is just 2 wt\%.\cite{liu2015new}

\begin{figure*}[!htb]
    \centering
    \includegraphics[width=16cm]{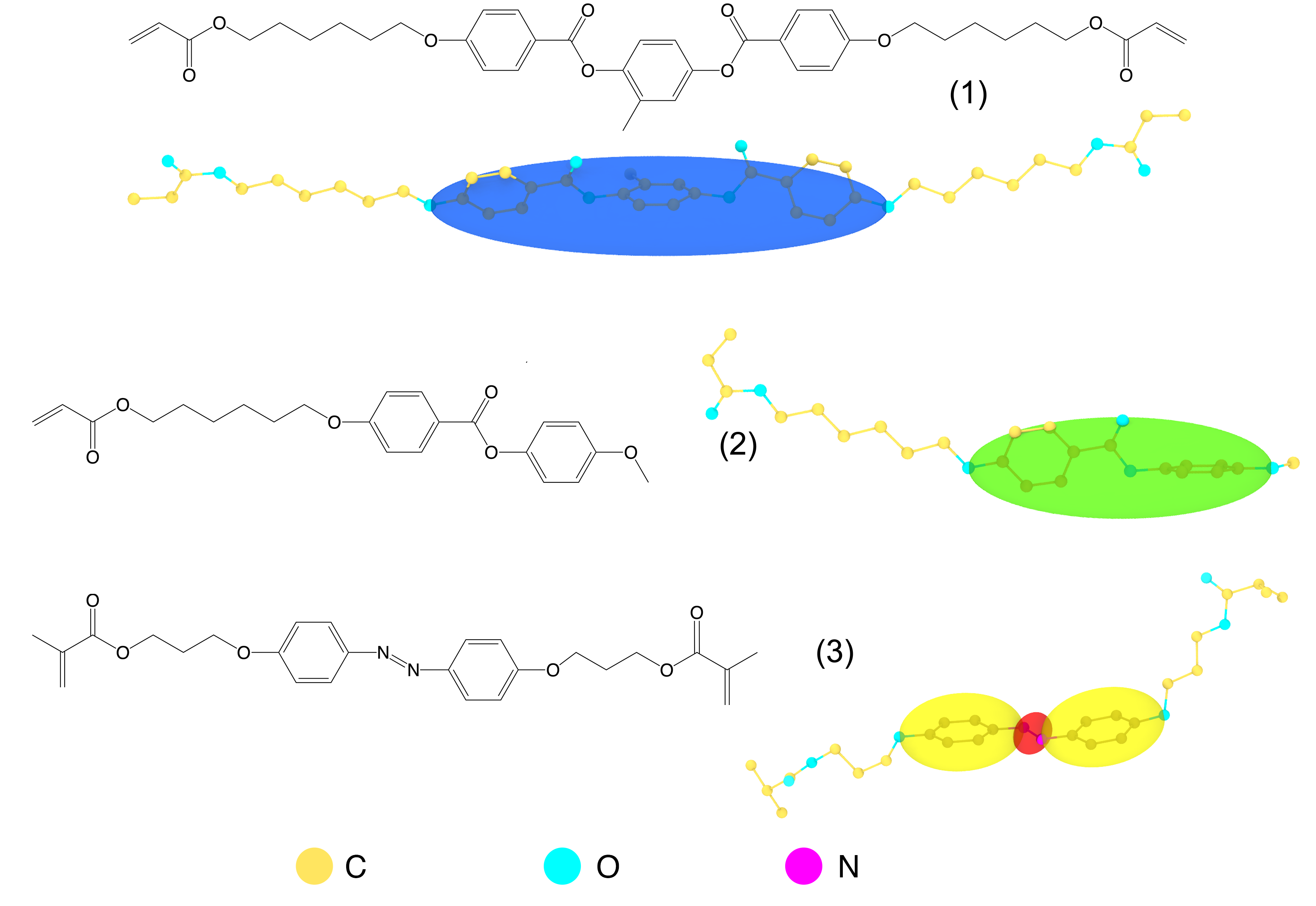}
    \caption{The chemical structure of the molecules (1) cross-linker (2) mono-acrylate, and (3) azo-diacrylate. The depiction of rod-like mesogens as ellipsoids, the atoms as spheres, and the bonds as lines for visualization of the ALCN unit cell. The Hydrogen atoms are omitted in the visualization. Note that the magenta-coloured N atoms are on either side of the red ellipsoid in between the two benzene rings in the azo-diacrylate (3).}
    \label{fig:Structures}
\end{figure*}
In this work, we have developed a computational framework using Materials Studio,\cite{materialstudio} LAMMPS\cite{lammps} and Python\cite{python} to simulate the continuous \trans-\cis-\trans~isomerization cycles of the azo-mesogen embedded in a polymer network at atomistic scale and study its influence on the distortion of the surrounding polymer network. The developed computational framework to simulate the dynamic isomerization of azobenzene embedded in LCNs is presented in the following section. In the later sections, we show that this dynamic isomerization is necessary for density reduction, and a mere statistical accumulation of \cis~isomers do not induce any changes to the density. We further use a probabilistic approach to simulate the influence of varying light intensities of the two wavelengths on the dynamic isomerization and the resulting density changes. Consequently, we demonstrate through molecular dynamic simulations the aforementioned counter-intuitive relationship between the density change and the intensities of the light sources. We then elucidate the underlying physics for such a counter-intuitive relation by analyzing the dihedral oscillation frequencies of the azo molecules.
\section{Computational Method}\label{Sec:Model}
\subsection{Preparation of ALCN computational sample}\label{sec:prep}
\begin{table*}[!htb]
\begin{ruledtabular}
\caption{The parameters of the dihedral energy of the C-N=N-C unit to simulate a photo-induced isomerization of the azo-mesogen.\cite{heinz2008photoisomerization}}
\label{Tab:dihedral_energy}
{\renewcommand{\arraystretch}{1.5}%
\begin{tabular}{ccccccc}
Energy & $K_1$ (\SI{}{\kilo\calorie\per\mol}{}) & $\phi_1$ (\SI{}{\degree}) & $K_2$ (\SI{}{\kilo\calorie\per\mol}{}) & $\phi_2$ (\SI{}{\degree}) & $K_3$ (\SI{}{\kilo\calorie\per\mol}{}) & $\phi_3$ (\SI{}{\degree})\\
\hline
$\Eth$ & 5.2 & 180 & 12 & 0 & 0 & 0\\
$\Etc$ & 34 & 0 & 5.2 & 180 & 0 & 0  \\
$\Ect$ & 34 & 180 & 0 & 0 & 0 & 0  \\
\end{tabular}
}
\end{ruledtabular}
\end{table*}
A cross-linked ALCN unit cell is created using Materials Studio.\cite{materialstudio} We selected a mixture of 10 cross-linker molecules (1), 100 mono-acrylate molecules (2) and 10 azo-diacrylate molecules (3) (corresponding to 8 mol\% azo-mesogens). The chemical structure of the above-mentioned molecules is presented in \cref{fig:Structures}. All 120 molecules were first packed in a cubic unit cell of size \SI{42.69}{\angstrom} with an average density of \SI{1.1}{\gram\per\cubic\centi\meter}. The mesogens' average alignment was along the unit cell's X-axis. This uncross-linked sample is then equilibrated using geometry optimization, followed by NVT and NPT dynamics for \SI{4}{\ns} and \SI{6}{\ns}, respectively. The equilibrated assembly is then cross-linked by artificially forming bonds between the reactive sites of the acrylate end in the monomers if they are within a set cutoff distance. The cutoff distance is initialized to be \SI{2.5}{\angstrom} and then increased in steps of \SI{0.2}{\angstrom} until 25\% of the reactive sites are polymerized. The polymerization procedure in this work is adapted for free-radical polymerization of acrylates from the existing methods in the literature.\cite{varshney2008molecular,choi2014photo,choi2016molecular} The polymerized assembly is then transferred from Materials Studio\cite{materialstudio} to LAMMPS\cite{lammps} and subjected to energy minimization, followed by NVE of \SI{1}{\ns} to initialize the velocities, NVT of \SI{1}{\ns} to reach a temperature of 300 K, NPT dynamics of \SI{1}{\ns} to initialize pressure to 1 atm and then NVT, NPT and NVE dynamics for \SI{3}{\ns}, \SI{10}{\ns} and \SI{1}{\ns}, respectively for complete equilibration. The box dimensions are allowed to change independently of each other. The time step for the equilibrium dynamics is \SI{1}{\fs} with a periodic boundary. The evolution of temperature, dimensions of the simulation domain, density and the scalar orientational order of the computational sample are presented in Fig. S2 during the equilibration. The final ALCN sample consists of 7278 atoms. The equilibrated sample had a density of \SI{1.12}{\gram\per\cubic\centi\meter}. The scalar orientational order parameter defined as $Q = \langle (3 \cos^2{\theta} - 1)/2 \rangle $ (with $\theta$ being the angle between the nematic axis and the mesogen) is 0.54 for the equilibrated ALCN sample. The potential energy of the constituent atoms is calculated using the Polymer Consistent Force-Field (PCFF).\cite{sun1994force} 
\subsection{Numerical implementation of photo-induced isomerization}
The photo-induced isomerization is simulated by modifying the parameters ($K_i, \phi_i, i=1,2,3$ in \cref{tors}) governing the torsion (dihedral terms) energy ($\Ed$) of the C-N=N-C unit of the azo-mesogen.\cite{heinz2008photoisomerization} 
\begin{multline}
    \Ed = K_1 (1 - \cos(\phi - \phi_1)) + K_2 (1 - \cos(2\phi - \phi_2)) \\ + K_3 (1 - \cos(3\phi - \phi_3)). \label{tors}
\end{multline}
%
Here $\phi$ is the C-N=N-C dihedral angle. The parameters that correspond to thermal equilibrium ($\alpha$ = thermal), \trans-to-\cis~($\alpha$ = $t \, \to \, c$) and \cis-to-\trans~($\alpha$ = $c \, \to \, t$) isomerizations are given in \cref{Tab:dihedral_energy}. The literature shows that the methodology of simulating the isomerization by modifying the dihedral energy parameters is successful in simulating different systems.\cite{tian2013reactive,choi2014photo,choi2016molecular} Even though the isomerization pathway induced by varying the dihedral energy parameters does not replicate the rotation or inversion-based isomerization mechanism of azobenzene,\cite{hugel2002single,fujino2002femtosecond,bandara2012photoisomerization} the current method seems able to simulate the essence of \trans-\cis-\trans~isomerization cycles. We consider the azo-molecule to be in \trans~state for $\phi > \SI{90}{\degree}$ and in \cis~state for $\phi < \SI{90}{\degree}$.

\subsubsection{Cyclic switching of isomerization}\label{sec:cyclic}
The ALCN computational sample is equilibrated in an NPT ensemble for a time interval of $\ts$ \SI{}{\ps} with the dihedral energy parameters ($K_i, \phi_i, i=1,2,3$ in \cref{tors}) of all the C-N=N-C units set to $\Etc$ and then equilibrated again for $\ts$ \SI{}{\ps} with the dihedral energy parameters of all the C-N=N-C units set to $\Ect$. The two equilibration steps are repeated till the total simulation time reaches \SI{100}{\ps}. The box shape is allowed to change during the NPT ensemble to accommodate the anisotropic changes to the unit cell dimensions. The time step for the equilibration is taken to be \SI{0.1}{\fs}. 

\subsubsection{Probabilistic switching of isomerization}\label{sec:prob}
\begin{figure}[!htb]
    \centering
    \includegraphics[width=8.7cm]{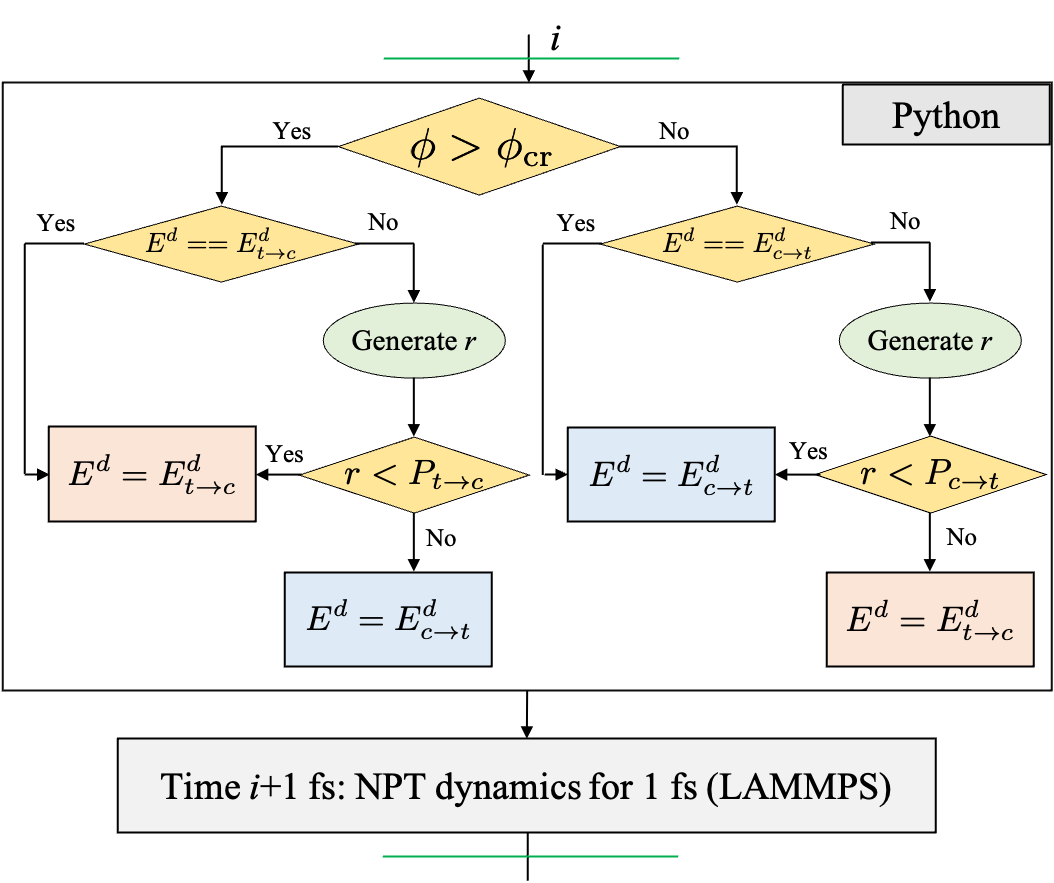}
    \caption{A flow chart illustrating the algorithm to simulate the influence of dynamic isomerization of azo-mesogens with different probabilities for \trans-to-\cis~($\Ptc$) and \cis-to-\trans~($\Pct$) reactions on the ALCN.}
    \label{fig:flow_chart}
\end{figure}

The flow chart shown in \cref{fig:flow_chart} illustrates the procedure to simulate the probabilistic isomerization. The dihedral energy parameters ($K_i, \phi_i, i=1,2,3$ in \cref{tors}) of the individual C-N=N-C units are updated based on the probabilities of \trans-to-\cis~($\Ptc$) and \cis-to-\trans~($\Pct$) isomerizations. The ALCN computational sample is equilibrated for \SI{1}{\fs} with a time step of \SI{0.01}{\fs} in an NPT ensemble allowing for changes to the shape of the simulation box with the set dihedral energy parameters. Parameters of $\Ed$ are updated at the beginning of every equilibration step with $\Phicri$ = \SI{90}{\degree} and $r$ being a random number between 0 and 1. 
\begin{figure*}[!htb]
    \centering
    \includegraphics[width=17.8 cm]{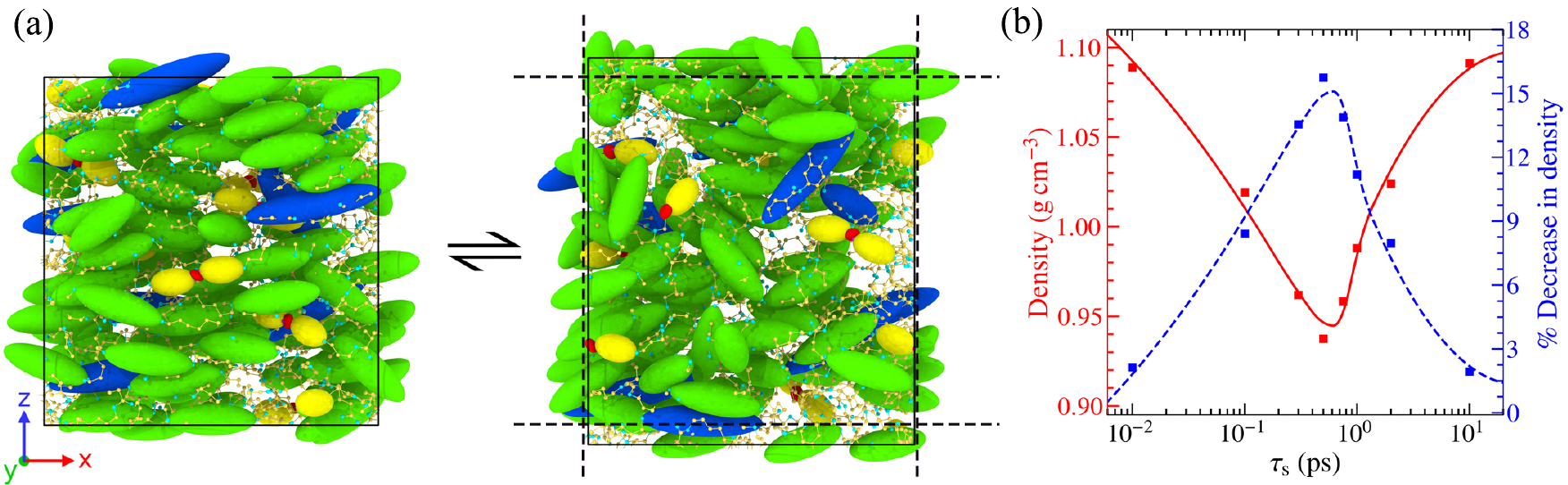}
    \caption{(a) The ALCN computational sample before the start (left) and after \SI{100}{\ps} (right) of cyclic switching between \trans~and \cis~isomer states with $\ts = \SI{0.5}{\ps}$ illustrating the distortion of the unit cell. The dashed lines in the right figure indicate the dimensions of the original unit cell (see \cref{fig:Structures} for details regarding the color coding and representation). The evolution of the ALCN sample during the first \SI{20}{\ps} of cyclic isomerization is also available as Movie S1. (b) The reduction in density due to the repeated switching of the dihedral energy parameters to induce \trans-to-\cis~and \cis-to-\trans~isomerizations at different time intervals ($\ts$). }
    \label{fig:Figure_DT}
\end{figure*}
\section{Results}\label{Sec:Results}
In order to verify the influence of isomerization on the polymer network, the parameters of the torsional energy ($\Ed$) of all the azo-molecules are modified to initiate \trans-to-\cis~isomerization, and then the ALCN unit cell is equilibrated for \SI{200}{\ps} in the NPT ensemble. The final system with 100\% \cis~population did not show any significant change in density (Fig. S3). However, a slight variation in the dimensions was observed. Since only 8 mol\% of azo-molecules are present, the isomerization from \trans~state to \cis~may not have provided enough network-pulling effect in the small computational sample to generate any noticeable changes to unit cell dimensions. Such an insignificant change to unit cell dimensions is in accordance with the literature\cite{choi2014photo} where isomerization of 100\% azo-system did not result in any density change, but a dimension change of 3\% was observed for 25\% \trans-to-\cis~conversion. The 100\% \cis~system is then converted to complete \trans~system and equilibrated again for \SI{200}{\ps}, where we again observe that the density changes to be insignificant. A time step of \SI{0.1}{\fs} is used for the equilibrium studies. The complete isomerization takes around \SI{0.2}{\ps}, and for the rest of the dynamics, the isomers stay in their respective state, resembling more of a static equilibrium than a dynamic equilibrium. 

\subsection{Dynamic \trans-\cis-\trans~isomerization cycles with different switch-time intervals}\label{Sec:Res_Cyclic}
The azo-mesogens in the computational ALCN sample are subjected to continuous switching between the two types of dihedral energy parameters that can induce \trans-to-\cis~and \cis-to-\trans~isomerizations to simulate a state of dynamic isomerization equilibrium. \update{Molecular dynamic simulations with various time intervals (switch-time interval, $\ts$) between each switching were performed (see Supplementary Note S2 and Fig. S4).} The computational unit cell visualized using OVITO\cite{stukowski2009visualization} before the start and after \SI{100}{\ps} of cyclic switching with $\ts = \SI{0.5}{\pico\second}$ is shown in \cref{fig:Figure_DT}a and \update{the evolution of the unit cell for the first \SI{20}{\ps} is also available as Movie S1.} The resultant density of the network as the azo-mesogens are repeatedly switched from \trans-to-\cis~and \cis-to-\trans~is plotted in \cref{fig:Figure_DT}b for various $\ts$, indicating that \SI{0.5}{\ps} is the optimal switch-time interval. The polymer network retains its reduced density as long as the isomers switch states and reverts to its original density (achieves equilibrium) as soon as the switching stops as seen in Fig. S5. \update{Therefore, it can be concluded that the repeated \trans-\cis-\trans~isomerization cycles at optimal time intervals cause the polymer network to be out of equilibrium, significantly reducing the density.}
Even though the relation between reduction in density with $\ts$ (\cref{fig:Figure_DT}b) from simulations and different ratios of intensities from experiments\cite{liu2015new}~(Fig. S1) are qualitatively similar, it is not clear if a direct correspondence can be drawn between varying intensities and switch-time intervals. Furthermore, in the simulations mentioned above, all azo-molecules are isomerized together from one state to another, which may not be true practically. For instance, the probability of isomerization strongly depends on the local intensities of the two wavelengths and other factors such as absorption coefficients and quantum efficiencies of isomerization, all of which induce some form of stochasticity to the whole process. We now simulate a more realistic system using probability-based isomerization criteria. 

\subsection{Dynamic \trans-\cis-\trans~isomerization cycles with probabilistic switching}\label{Sec:Res_Probs}
\begin{figure*}[!htb]
    \centering
    \includegraphics[width=17.8 cm]{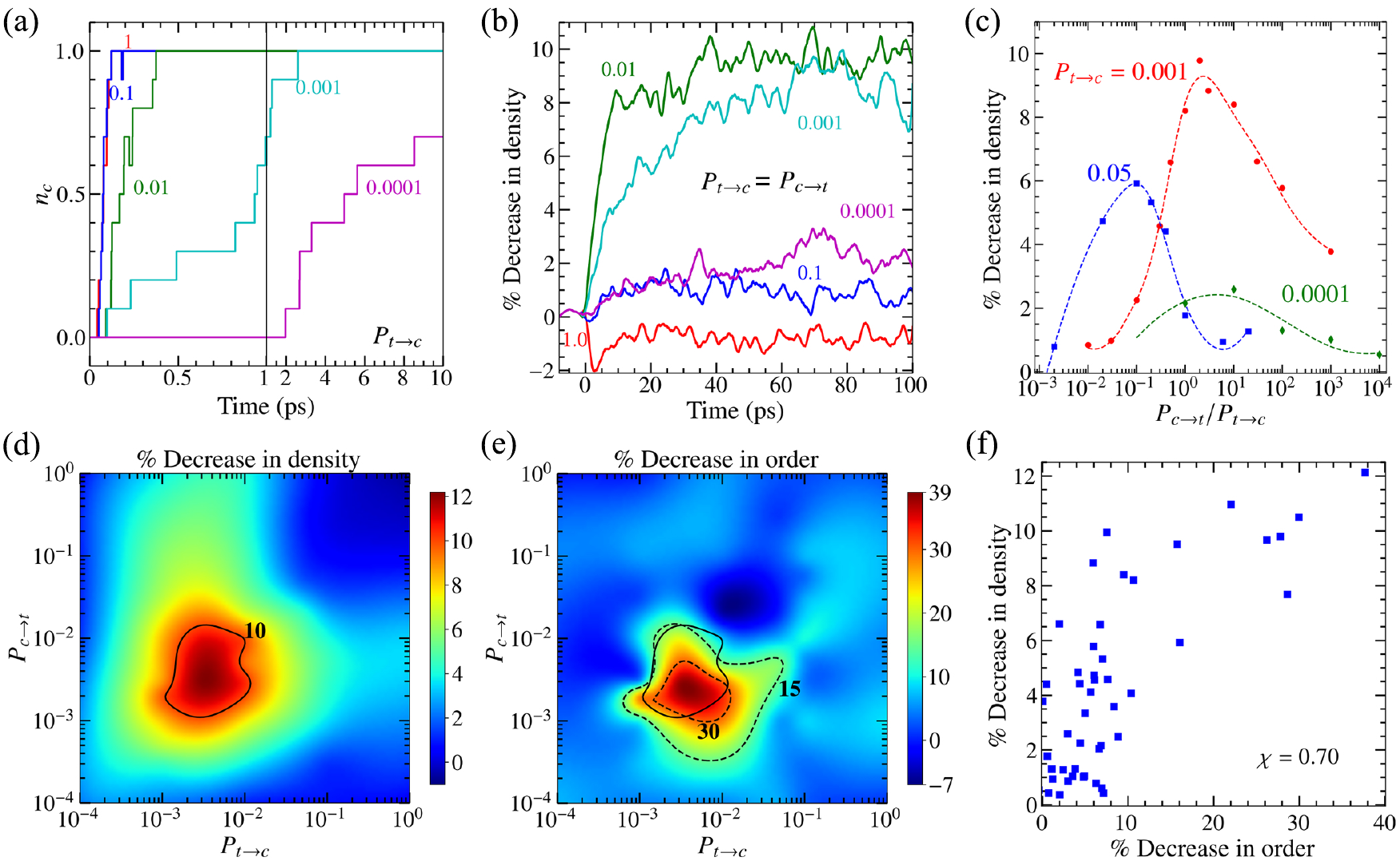}
    \caption{(a) The evolution of $\nc$ with time for different levels of $\Ptc$ with $\Pct$ = 0. (b) The evolution of density for various levels of equal \trans-to-\cis~and \cis-to-\trans~isomerization probabilities~($\Ptc = \Pct$). (c) \update{The decrease in density as a function of the probability ratio, $\Pct/\Ptc$ for different values of $\Ptc$}. The contour plot of (d) decrease in density and (e) decrease in order as a function of $\Ptc$ and $\Pct$. (f) The decrease in density plotted as a function of decrease in order to illustrate their strength of correlation ($\chi$). }
    \label{fig:Figure_Prob}
\end{figure*}
The probability of isomerization decides the time taken for complete isomerization as seen in \cref{fig:Figure_Prob}a where the evolution of $\nc$ is shown with different values for \trans-to-\cis~isomerization probability ($\Ptc$) while keeping \cis-to-\trans~isomerization probability ($\Pct$) = 0. The effects of changing probability parallels that of varying intensities in this scenario. Such a correspondence can be drawn because increasing $\Ptc$ (simulation) and intensity (experiments) have the same effect of reducing the time for isomerization. The evolution of density during the dynamic isomerization for equal probabilities of \trans-to-\cis~isomerization and \cis-to-\trans~isomerization (i.e., $\Ptc = \Pct$) is plotted in \cref{fig:Figure_Prob}b. For the case of $\Ptc = \Pct = 1$ or 0.1, the azo-molecules just oscillate around an equilibrium value of the C-N=N-C dihedral angle ($\Phicnnc \approx \SI{90}{\degree}$, see Fig. S6a) without causing much distortion of the polymer network and hence does not alter the density. At very low values of $\Ptc = \Pct = 0.0001$, the azo-mesogens isomerize rarely so that the average $\Phicnnc$ changes so slowly such that the density changes are not significant (see Fig. S6b). When $\Ptc = \Pct$ = 0.01 or 0.001, the azo-mesogens isomerize from \trans-to-\cis~and \cis-to-\trans~at close intervals causing large distortions to the polymer network (see Figs. S6c and d). Moreover, as the dynamic isomerization continues, the polymer network is continuously disturbed, leading to significant density changes ($\approx 9\%$). \update{\Cref{fig:Figure_Prob}c shows the decrease in density as a function of the probability ratio, $\Pct/\Ptc$ for different values of $\Ptc$. When the value of $\Pct$ is much lower than $\Ptc$, the isomers always stay in \cis~state without any large oscillations leading to lower density changes (see Fig. S7a for the case of $\Ptc = 0.001$ and $\Pct/\Ptc = 0.01$). In contrast, when the value of $\Pct$ is much greater than $\Ptc$, the azo-molecules get pushed to \trans~state as soon as they isomerize to \cis~state, leading to low magnitude oscillations (see Fig. S7b for the case of $\Ptc = 0.001$ and $\Pct/\Ptc = 1000$) and hence, low changes to the density. At an optimal value of $\Pct$ for a given $\Ptc$, the isomers switch at an optimal rate to significantly distort the polymer network and keep it from reaching equilibrium leading to a maximum reduction in density. The cases of $\Ptc = 0.05$ and 0.0001 also show a similar inverse `U' relationship between density decrease and probability ratios. Such a non-monotonous relationship between the probabilities and density reduction is similar to the experimental observations\cite{liu2015new} where a maximum reduction in density is observed for a specific ratio of the intensities. At a given $\Ptc$, the maximum density reduction possible for various $\Pct$ also differs as seen from \cref{fig:Figure_Prob}c and d. Such a behaviour is analogous to the maximum density reduction being lower for $\Iuv$ = \SI{100}{\unitsI} than $\Iuv$ = \SI{300}{\unitsI} in the experiments\cite{liu2015new} (see Fig. S1). Moreover, the values of $\Pct/\Ptc$ at which the density decrease is maximum for a given $\Ptc$ also changes as seen in \cref{fig:Figure_Prob}c. The highest density reduction observed through the simulations is 12\% for the case of $\Ptc = \Pct$ = 0.003 (see \cref{fig:Figure_Prob}d). Subsequently, the simulation results indicate that probabilities (intensities) higher than the optimal value could result in lower-density changes. The above results present a satisfactory numerical evidence to the hypothesis that the dynamic isomerization of azo-molecules induces density reduction of an ALCN and there exists an optimal combination of probabilities (intensities) that generate a maximum density reduction.} 

\begin{figure*}[!htb]
    \centering
    \includegraphics[width=17.80 cm]{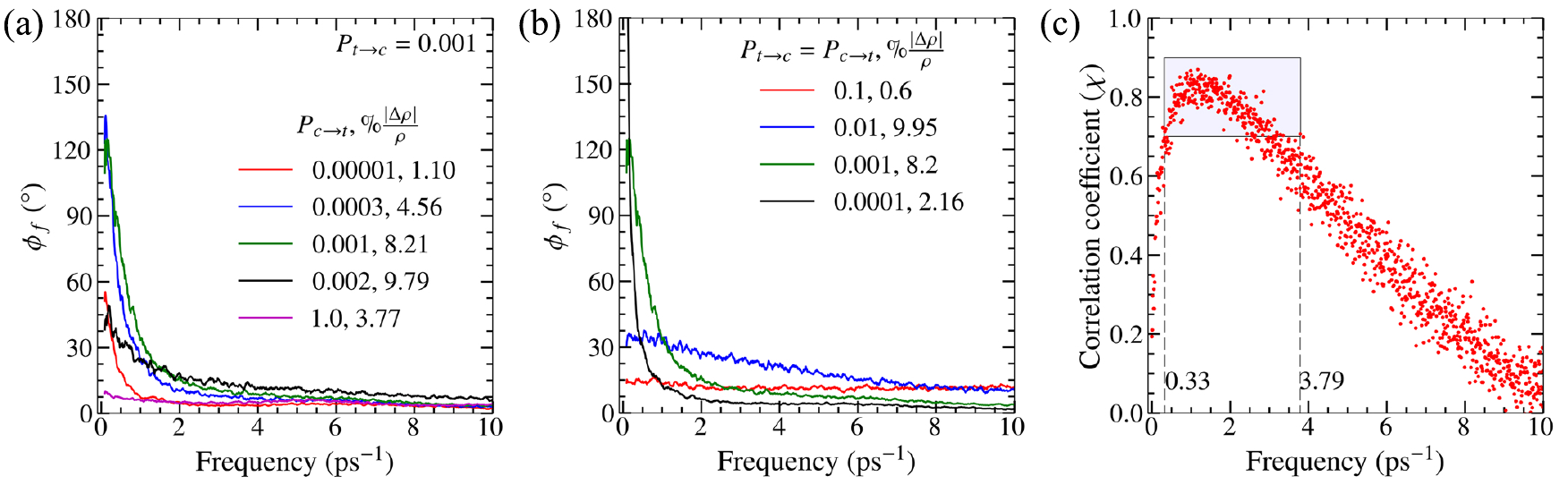}
    \caption{(a,b) The average of the FFTs of the dihedral angles of the individual azo-mesogens at a given frequency for different probabilities of \trans-to-\cis~and \cis-to-\trans~isomerizations and the corresponding percentage change in density ($\% \frac{\Delta \rho}{\rho}$). (c) The correlation coefficient between the average of the amplitude of oscillations of the dihedral angles at a given frequency obtained through FFT with the density decrease indicating that the dihedral oscillations at frequencies in the range \SI{0.33}{\per\ps} to \SI{3.79}{\per\ps} influence density strongly.}
    \label{fig:Figures_Prob2a}
\end{figure*}
The decrease in the scalar orientational order parameter ($Q$) after \SI{100}{\ps} of dynamic isomerization with different \trans-to-\cis~and \cis-to-\trans~probability combinations is plotted in \cref{fig:Figure_Prob}e. It can be seen that the probabilities $\Ptc \approx 0.003,\, \Pct \approx 0.003$ corresponding to maximum density reduction also show a maximum decrease in order. However, the dependence of density and order on $\Ptc, \Pct$ are different and are not correlated linearly. Specifically, a density reduction of 10\% occurs for order reduction of 15\% as well as 30\%. The Pearson correlation coefficient between the density and order was 0.70 (see \cref{fig:Figure_Prob}f). The reduction in density and order is due to dynamic \trans-\cis-\trans~isomerization cycles, and the order change may not be the primary cause for density change. The atomistic scale simulations in the literature\cite{choi2014photo} also indicate that order change may not directly result in density change. In particular, they simulated isomerization of a 100\% azo-system which showed a significant decrease in order while the density stayed constant. 
\section{Discussion}\label{Sec:Discussion}
\begin{figure}[!htb]
    \centering
    \includegraphics[width=7.0 cm]{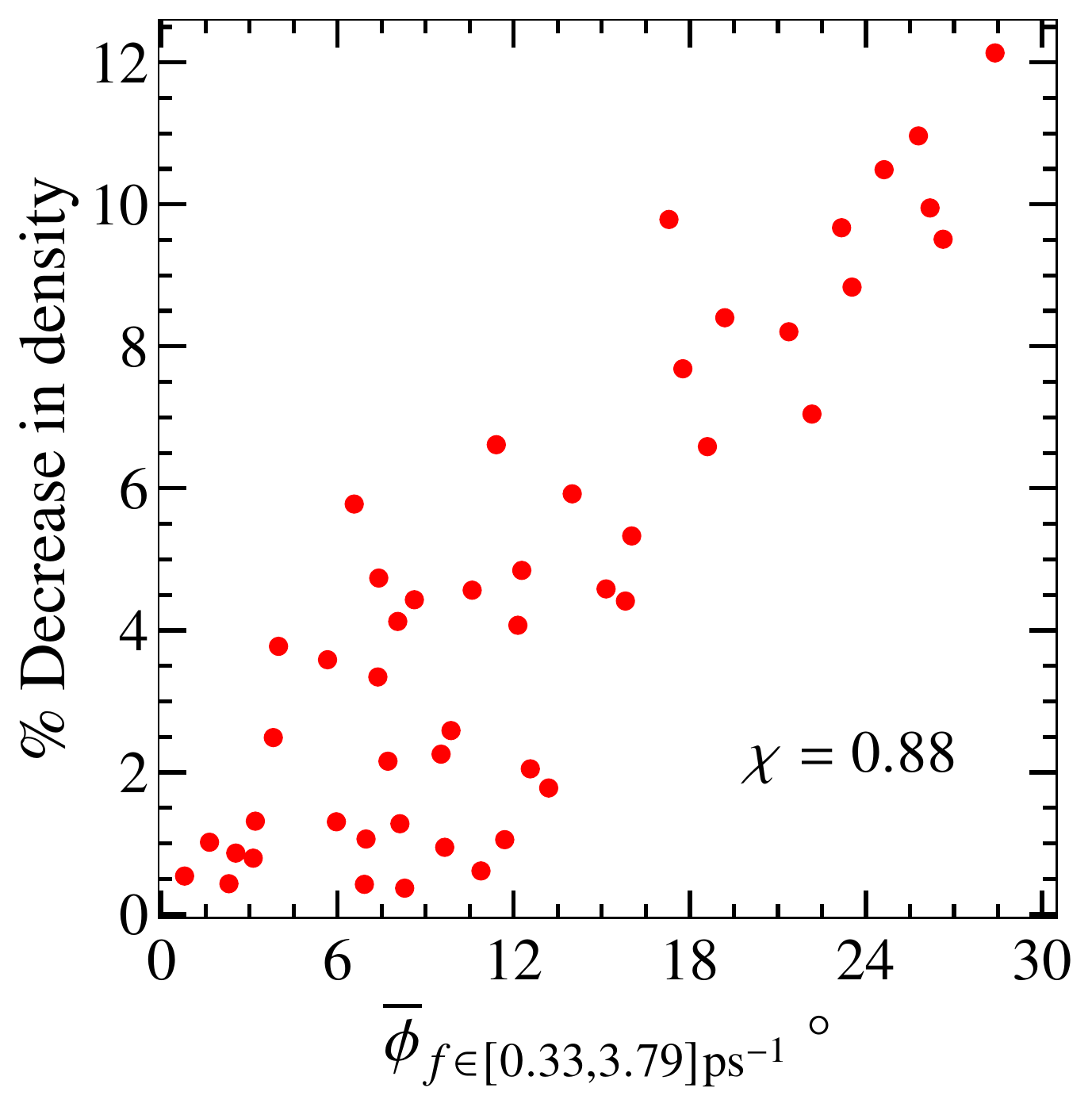}
    \caption{The decrease in density as a function of average amplitudes of oscillations of the C-N=N-C dihedral angles in the frequency range \SI{0.33}{\per\ps} to \SI{3.79}{\per\ps} indicating a strong correlation ($\chi$ = 0.88).}
    \label{fig:Density_FR}
\end{figure}

%
From the results presented, it is shown that the variations of the C-N=N-C dihedral angle ($\Phicnnc$) of the azo-mesogens in the network due to switching of isomerization states result in density changes. Now, to determine the nature of variation in $\phi$ that results in significant changes in density at only specific probabilities, we use the Pearson correlation coefficient between different characteristics of $\phi$ and density change. {The temporal variations of the dihedral angle for one of the azo molecules of the ALCN during dynamic isomerization for different probability cases are shown in Figures S8 and S9.} The average of the Fast Fourier Transform (FFT) of the dihedral angles of the individual azo-molecules at a given frequency ($\phi_f$) for different probabilities are presented in \cref{fig:Figures_Prob2a}a and \cref{fig:Figures_Prob2a}b. A higher amplitude of the FFT at a given frequency may be expected to result in higher density change. But, consider the cases of $\Ptc=\Pct = $ 0.01 or 0.001 (\cref{fig:Figures_Prob2a}b); their FFTs intersect at around \SI{1}{\per\ps} although they have similar density reduction. Hence, oscillations at any frequency do not contribute equally to the density change, and specific frequencies may be more optimal. The existence of an optimal frequency for maximal density reduction has also been observed for the AC electric field responsive LCNs\cite{liu2017protruding}. To determine the optimal frequencies of oscillations of the dihedrals in photo-responsive ALCNs, we have correlated the amplitude of oscillations at each frequency ($\phi_f$) with the decrease in density as shown in \cref{fig:Figures_Prob2a}c. The correlation coefficient ($\chi$) is greater than 0.7 for frequencies in the range \SI{0.33}{\per\ps} to \SI{3.79}{\per\ps}, indicating that the oscillations at these frequencies contribute significantly to the density reduction. The decrease in density and the average of the amplitude of oscillations in the frequency range from \SI{0.33}{\per\ps} to \SI{3.79}{\per\ps} have a correlation of 0.88, indicating the presence of strong relationship (see \cref{fig:Density_FR}). 
The mean of the frequencies which strongly correlate with density reduction is $\approx$ \SI{2}{\per\ps} corresponding to a time scale of \SI{0.5}{\ps} indicating that the switch-time of around \SI{0.5}{\ps} contributes to the maximal density reduction. Note that a maximum density change of 15.7\% was observed for a switch-time interval of \SI{0.5}{\ps} in our earlier discussion (see \cref{fig:Figure_DT}c) where all the dihedrals are switched between \trans~and \cis~states at specific time intervals. Therefore, we propose that the dynamic isomerizations that occur with a time period of \SI{0.5}{\ps} are optimal for distorting the polymer network and keeping it from attaining an equilibrium resulting in significant density reduction. 
\section{Summary}\label{Sec:Summary}
In summary, for the first time, we have validated the hypothesis that dynamic isomerization induces density changes to the azo-modified liquid crystal polymer network through atomistic scale simulations. The cyclic switching of all azo-molecules between \trans~and \cis~states resulted in a density change of 15.7\%, despite azo-molecules making up only 8 mol\% of the computational ALCN sample. We have also explained the physics governing the experimental observation of the existence of light intensities for which density changes are a maximum.\cite{liu2015new} This explanation is brought about by drawing an analogy between intensity ratios (in experiments) and probability ratios (in simulations) of isomerization. Lastly, we have identified the frequencies of oscillations of the C-N=N-C dihedrals which contribute significantly to the density changes. The proposed methodology can be extended in future to understand the influence of the dynamic isomerization cycles on the glass transition temperature and the light-induced softening of ALCNs, which could then provide better insights into the origin of chaotic self-oscillations.\cite{kumar2016chaotic} The network pulling effect based mechanism generated by accumulation of \cis~isomers typically results in approximately 2\% deformation,\cite{van2007glassy,harris2005large} while, the dynamic isomerization can yield deformations as high as 10\%.\cite{liu2015new} The applications of liquid crystal polymers in soft grippers,\cite{liu2014self,lyu2022robotic} active control of friction,\cite{liu2014light,liu2015reverse} cell adhesion and migration\cite{koccer2017light} requiring large topological changes may be induced through dynamic isomerization. Consequently, the improved understanding of dynamic isomerization and the resulting density changes in ALCNs, as provided in this work, can aid the design of the coatings for such applications.

\section*{Supplementary Material}
\begin{itemize}
    \item Figures S1-S9: Experimentally observed dual-wavelength light induced density decrease; the evolution of temperature, unit cell dimensions, density and orientational order during equilibration of ALCN computational sample; the influence of static isomerization on the density; the influence of cyclic isomerization with different switch-time intervals on the density; the reversibility of the density reduction; the average of the dihedral angles of the azo-mesogens during probabilistic isomerization; the temporal variations of the dihedral angle during probabilistic isomerization. Note S1: A photo-physical model to estimate the fraction of isomers undergoing dynamic isomerization under dual-wavelength illumination. Note S2: The evolution of density during the cyclic switching of dihedral energy parameters with different switch-time intervals (pdf)
    \item The evolution of the computational sample during the repeated, continuous switching of all the azo-isomers between \trans~states and \cis~states with a switch-time interval of \SI{0.5}{\ps} for a duration of first \SI{20}{\ps}. The unit cell's boundaries before switching isomerization starts are highlighted as red lines. The mesogens are depicted as ellipsoids, the atoms as spheres, and the bonds as lines for visualization of the ALCN unit cell, as shown in Fig. 2. (Movie S1) (mp4)
\end{itemize}

\begin{acknowledgements}
The authors acknowledge the fruitful discussion with Prof. Dirk Broer from the Eindhoven University of Technology and the email conversation with Prof. Hendrik Heinz from the University of Colorado, Boulder, that proved essential for this work. The authors acknowledge the generous financial support from the Indian Institute of Technology Madras under the Institutes of Eminence (IoE) scheme funded by the Ministry of Education, Government of India. A.R. Peeketi acknowledges the financial support through Prime Minister’s Research Fellowship for conducting doctoral research at IIT Madras. 
\end{acknowledgements}

\section*{Data Availability Statement}
The data that support the findings of this study are available within the article and its supplementary material. The input scripts, potential files, and Python codes employed for the molecular dynamic simulations of the dynamic isomerization of azo-molecules embedded in a liquid crystal polymer network are made available at the GitHub repository, \\ \href{https://github.com/ARPeeketi/MD_ALCN_Dynamic_Isomerization}{https://github.com/ARPeeketi/MD\_ALCN\_Dynamic\_\\Isomerization}.

\section*{Conflict of Interest Statement}
The authors have no conflicts to disclose.
\bibliography{references_list}

\end{document}